\documentclass[preprint]{elsarticle}
\usepackage{amssymb,amsmath,graphicx,natbib,upgreek}
\begin{document}

\begin{frontmatter}

\title{Numerical Simulations of Mach Stem Formation via Intersecting Bow Shocks}
\author[ur]{E.C. ~Hansen\corref{cor1}}
\ead{ehansen@pas.rochester.edu}
\author[ur]{A. ~Frank}
\author[rice]{P. ~Hartigan}
\author[lanl]{K. ~Yirak}
\cortext[cor1]{Corresponding author}
\address[ur]{Department of Physics and Astronomy, University of Rochester, Rochester, NY 14627-0171, USA}
\address[rice]{Department of Physics and Astronomy, Rice University, 6100 S. Main, Houston, TX 77521-1892, USA}
\address[lanl]{Los Alamos National Laboratory, Los Alamos, NM 87545, USA}

\begin{abstract}
Hubble Space Telescope observations show bright knots of H$\alpha$ emission within outflowing young stellar jets.
Velocity variations in the flow create secondary bow shocks that may intersect and lead to enhanced emission.
When the bow shocks intersect at or above a certain critical angle, a planar shock called a Mach stem is formed.
These shocks could produce brighter H$\alpha$ emission since the incoming flow to the Mach stem is parallel to the shock normal.
In this paper we report first results of a study using 2-D numerical simulations designed to explore Mach stem formation at the intersection of bow shocks formed by hypersonic ``bullets'' or ``clumps''.
Our 2-D simulations show how the bow shock shapes and intersection angles change as the adiabatic index $\gamma$ changes.
We show that the formation or lack of a Mach stem in our simulations is consistent with the steady-state Mach stem formation theory.
Our ultimate goal, which is part of an ongoing research effort, is to characterize the physical and observational consequences of bow shock intersections including the formation of Mach stems.
\end{abstract}


\end{frontmatter}

\section{Introduction}
\label{sec:intro}
Astrophysical jets are heterogeneous beams of plasma traveling at supersonic velocities.
They can be found in a variety of environments at different scales (see \cite{Frank14} for an overview).
Herbig-Haro (HH) jets are associated with young stellar objects (YSOs) and the star formation process.
They propagate away from YSOs and interact with the interstellar medium.
HH jets are ubiquitous in star forming regions because they are, most likely, closely correlated to the accretion processes creating those stars.
Thus it follows that time variability in the accreting disk produces variability in the outflowing jet \cite{Cabrit90,Hartigan95}.

Structures within jet beams can be caused by variations in the momentum injection at the jet source \cite{Raga90}.
This variability results in a ``clumpy'' jet flow leading to a variety of heterogeneous interactions along the jet beam (see e.g. \cite{Raga02ApJ,Hartigan05,Yirak12}).
Hubble Space Telescope (HST) time-series observations of HH objects reveal localized bright knots in H$\alpha$ and regions of strong [S II] \cite{Hartigan11}.
Some of these H$\alpha$ features represent shock fronts caused by variable velocities, and the [S II] regions represent cooling regions behind shocks \cite{Heathcote96}.
Many groups have studied jet models with a variable injection velocity (e.g. \cite{Kajdic06,deColle06}).
Recent high resolution MHD simulations by Hansen et al. \cite{Hansen14} explored how these variable velocity jet models produce internal shocks and the H$\alpha$ and [S II] emission patterns they produce.

The HST observations also reveal that some of these H$\alpha$ knots are brighter than expected, and these are located at the intersection points between separate bow shocks.
There is still some uncertainty as to why these knots are brighter, but one possible explanation is that a Mach stem formed at the intersection.
When bow shocks intersect at an angle at or above a certain critical value, a third shock (Mach stem) will form.
Mach stems form perpendicular to the direction of flow, so incoming plasma will encounter a planar shock instead of an oblique one.
A planar shock would theoretically lead to brighter emission at this location.
In \cite{Hartigan11}, Figure 7 shows HST images of a region of HH 2 with H$\alpha$ in green and [S II] in red.
This region is an ideal laboratory for studying potential Mach stems as it consists of many clumps and small secondary bow shocks.

High Energy Density Laboratory Experiments have also been conducted in order to understand Mach stems  \cite{Foster10,Yirak13}.
These experients were also important to the efforts surrounding inertial-confinement fusion because they are thought to be related to various plasma instabilities such as the Kelvin-Helmholtz instability \cite{Goldman99}.
By creating oblique shocks running past a reflecting wall these experiments created situations that are analogous to the astrophysical scenario of two intersecting bow shocks.
The main goal of these experiments was to explore the Mach stem growth rate as a function of angle between the shock flow and the obstructing wall.

In this work, we focus on the astrophysical scenario by conducting a set of high resolution simulations of intersecting bow shocks in 2-D.
We explore how the ratio of specific heats $\gamma$ affects the shape of the bow shock and whether or not this allows a Mach stem to form given the separation $d$ of clumps producing the bow shocks.
We show that our simulations are in good agreement with the theory of Mach stem formation, and in future work, we will explore other properties and consequences of Mach stems, such as emission, more thoroughly.

The structure of the paper is as follows: in Section~\ref{sec:theory}, we give a brief overview of the theory of Mach stem formation.
In Section~\ref{sec:sims}, we present some of our numerical methods as well as our initial conditions for the simulations.
Section~\ref{sec:results} contains our simulation results and our interpretations.
Finally, we finish the paper with our conclusions in Section~\ref{sec:conc}.

\section{Theory}
\label{sec:theory}
\begin{figure}
\includegraphics[width=\linewidth]{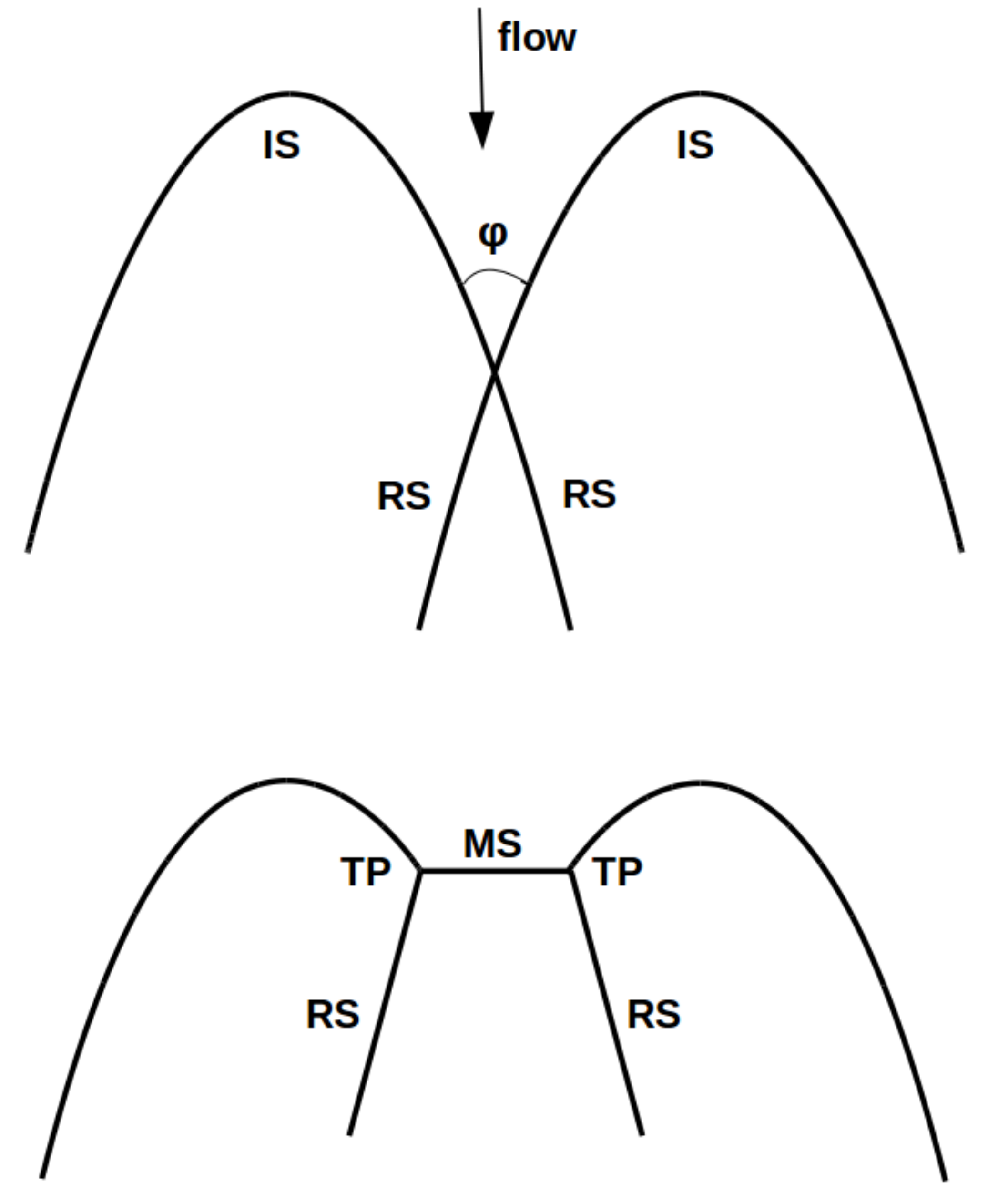}
\caption{Diagram of intersecting bow shocks with (\emph{bottom}) and without (\emph{top}) a Mach stem.
The acronyms are as follows: ``IS'' = incident shock (bow shock), ``RS'' = reflected shock, ``MS'' = Mach stem, ``TP'' = triple point.
Also shown is the direction of the flow and the included angle $\upvarphi$.}
\label{fig:reflect}
\end{figure}

A description of Mach stem formation, also known as Mach reflection, can be found in several texts (e.g. \cite{BenDor07,Courant99,Landau87}), so it will suffice to give a brief explanation here.
When two bow shocks intersect, they reflect off of each other forming symmetric reflected oblique shocks; this is known as regular reflection.
Without a Mach stem, a gas parcel passes through the incident bow shock and then through the reflected shock.
The gas is deflected by the shocks in such a way that its final velocity is in the same direction as its initial pre-shock velocity.
Once the intersection of the bow shocks reaches a critical angle, determined by the jump conditions for the colliding shocks, a gas parcel can no longer pass through both shocks and maintain its original direction of flow.
Therefore, a planar shock known as a Mach stem is formed so that the gas can continue to flow downstream in the same direction as the initial pre-shock flow.
The Mach stem extends from each bow shock at positions known as triple points which represent the new positions where the bow shocks reflect (see Figure~\ref{fig:reflect}).

For strong shocks, the critical angle $\upvarphi_c$ is only dependent on the ratio of specific heats $\gamma$.
A simple derivation of $\upvarphi_c$ leads to the following approximate formula \cite{Courant99}:

\begin{equation}
\upvarphi_c = 2\arcsin(\frac{1}{\gamma}) .\ \label{1}
\end{equation}

This approximation breaks down as $\gamma$ decreases, and it should be noted that the bow shock may be unstable to the Vishniac instability for $\gamma < 1.2$ \cite{Vishniac83}.
A more detailed derivation of this critical angle was done by De Rosa et al. \cite{deRosa92} and gives a more accurate equation.

\begin{equation}
\upvarphi_c = 2\arctan{\bigg[\frac{1}{\gamma-1}\sqrt{\frac{\sqrt{\gamma^2-1}}{\gamma-\sqrt{\gamma^2-1}}}\Big(1-\sqrt{\gamma^2-\gamma\sqrt{\gamma^2-1}}\Big)\bigg]} .\ \label{2}
\end{equation}

The values of $\gamma$ used in our simulations result in critical angles shown in Table~\ref{tab:phicrits} (in degrees).
The critical angle is a \emph{minimum}, so an intersection that occurs at or above this angle should form a Mach stem.
Note that as $\gamma$ decreases, $\upvarphi_c$ increases.
As $\gamma$ approaches 1, in other words as the gas becomes isothermal, the Mach stem should not form.

\begin{table}
\begin{center}
\begin{tabular}{|c|c|}
\hline
$\gamma$ & $\upvarphi_c$ \\
\hline
5/3 & 74.75 \\
\hline
1.4 & 83.31 \\
\hline
1.2 & 95.34 \\
\hline
1.01 & 137.46 \\
\hline
\end{tabular}
\end{center}
\caption{Critical angles $\upvarphi_c$ (in degrees) for selected values of $\gamma$.}
\label{tab:phicrits}
\end{table}

\section{Numerical Simulations}
\label{sec:sims}
\subsection{Methods}
\label{subsec:methods}
The simulations were carried out using AstroBEAR, a highly parallelized adaptive mesh refinement (AMR) multi-physics code.
See \cite{Cunningham09,Carroll12} for a detailed explanation of how AMR is implemented.
More details of the code can also be found at http://bearclaw.pas.rochester.edu/trac/astrobear.
Here we provide a brief overview of the physics implemented for this study.
The code solves the 2-D Euler equations of fluid dynamics:

\begin{subequations}\label{group3}
\begin{gather}
\frac{\partial \rho}{\partial t} + \boldsymbol{\nabla} \cdot \rho \boldsymbol{v} = 0 ,\ \label{3a}\\[\jot]
\frac{\partial \rho \boldsymbol{v}}{\partial t} + \boldsymbol{\nabla} \cdot (\rho \boldsymbol{v} \boldsymbol{v} + P\boldsymbol{I}) = 0 ,\ \label{3b}\\[\jot]
\frac{\partial E}{\partial t} + \boldsymbol{\nabla} \cdot ((E + P) \boldsymbol{v}) = 0 ,\ \label{3c}
\end{gather}
\end{subequations}
where $\rho$ is the mass density, $\boldsymbol{v}$ is the velocity, $P$ is the thermal pressure, $\boldsymbol{I}$ is the identity matrix, and $E$ is the total energy such that $E = \frac{1}{\gamma - 1} P + \frac{1}{2}\rho v^2$ (with $\gamma = \frac{5}{3}$ for an ideal gas).
The equations above represent the conservation of mass \eqref{3a}, momentum \eqref{3b}, and energy \eqref{3c}.

\subsection{Initial Conditions}
\label{subsec:init}
The simulations consist of 2 stationary clumps in an ambient medium with a wind sweeping over them from the top boundary.
In 2-D, these clumps are cross sections of cylinders.
This wind creates 2 intersecting comoving bow shocks which are ideal for studying Mach reflection.

All simulations use 4 levels of AMR which leads to an effective resolution of 160 cells per clump radius.
One clump radius is equal to 10 AU.
All boundaries use open boundary conditions except for the top which has the injected wind.
	
The ambient density and temperature are set at 10\textsuperscript{3} cm\textsuperscript{-3} and 10\textsuperscript{4} K respectively.
The clumps have a lower initial temperature of 10\textsuperscript{3} K and are ``over-dense'' at 10\textsuperscript{7} cm\textsuperscript{-3}.
This high clump density is used to keep the transmitted shocks propagating through the clumps at a low velocity.
Thus the clumps evolve slowly and do not affect the behavior of the bow shocks.
The wind has a density and temperature of 5 x 10\textsuperscript{3} cm\textsuperscript{-3} and 10\textsuperscript{4} K respectively.
It is injected from the top boundary at 50 km/s resulting in a Mach number of approximately 5.

We have run 16 different models with varying values for $\gamma$ and separation distance $d$.
The values for $\gamma$ that we tested are 5/3, 1.4, 1.2, and 1.01, and the values for $d$ that we tested are 20, 15, 10, and 5 in units of $r_{clump}$.
All simulations ran for approximately 75 years.

The aforementioned densities, temperatures, velocity, and length scale are within the commonly accepted ranges for HII regions within the interstellar medium (see \cite{Frank14}, and refs. therein).

\section{Simulation Results}
\label{sec:results}
The shapes of the simulated bow shocks were determined using curve fitting techniques.
This made it possible to calculate the intersection angles for clumps separated by different distances.
Thus using the fitted curves, one can calculate at what position the intersection angle becomes supercritical which can then be interpreted as a critical separation distance at which Mach stem formation becomes possible.

Unlike the critical angle $\upvarphi_c$, the critical distance $d_c$ is a \emph{maximum}; a separation distance at or below $d_c$ should form a Mach stem.
Table~\ref{tab:dcrits} gives the calculated values for $d_c$ for the values of $\gamma$ used in our simulations.
The trends seen in both Table~\ref{tab:phicrits} and Table~\ref{tab:dcrits} are consistent with each other.
As $\gamma$ decreases, $\upvarphi_c$ increases which implies that the bow shocks have to be closer to form a Mach stem.
Also, the bow shocks become narrower as $\gamma$ decreases which futher necesitates that the bow shocks be closer to form a Mach stem.
The bow shocks become narrower because the gas approaches isothermal conditions as $\gamma$ approaches 1, and the bow shock collapses as it becomes isothermal and loses pressure support.

\begin{table}
\begin{center}
\begin{tabular}{|c|c|}
\hline
$\gamma$ & $d_c$ \\
\hline
5/3 & 12.5 \\
\hline
1.4 & 6.2 \\
\hline
1.2 & 4.6 \\
\hline
1.01 & unstable \\
\hline
\end{tabular}
\end{center}
\caption{Critical separation distances $d_c$ (in units of $r_{clump}$) for selected values of $\gamma$.}
\label{tab:dcrits}
\end{table}

As was noted previously in Section~\ref{sec:theory}, shocks in flows with $\gamma < 1.2$ will be susceptible to the Vishniac instability, and our simulations confirm this.
The instability of the $\gamma = 1.01$ bow shock makes it impossible to determine its shape using curve fitting techniques.
Hence, the critical separation distance for $\gamma = 1.01$ cannot be calculated.

\begin{figure}
\includegraphics[width=\linewidth]{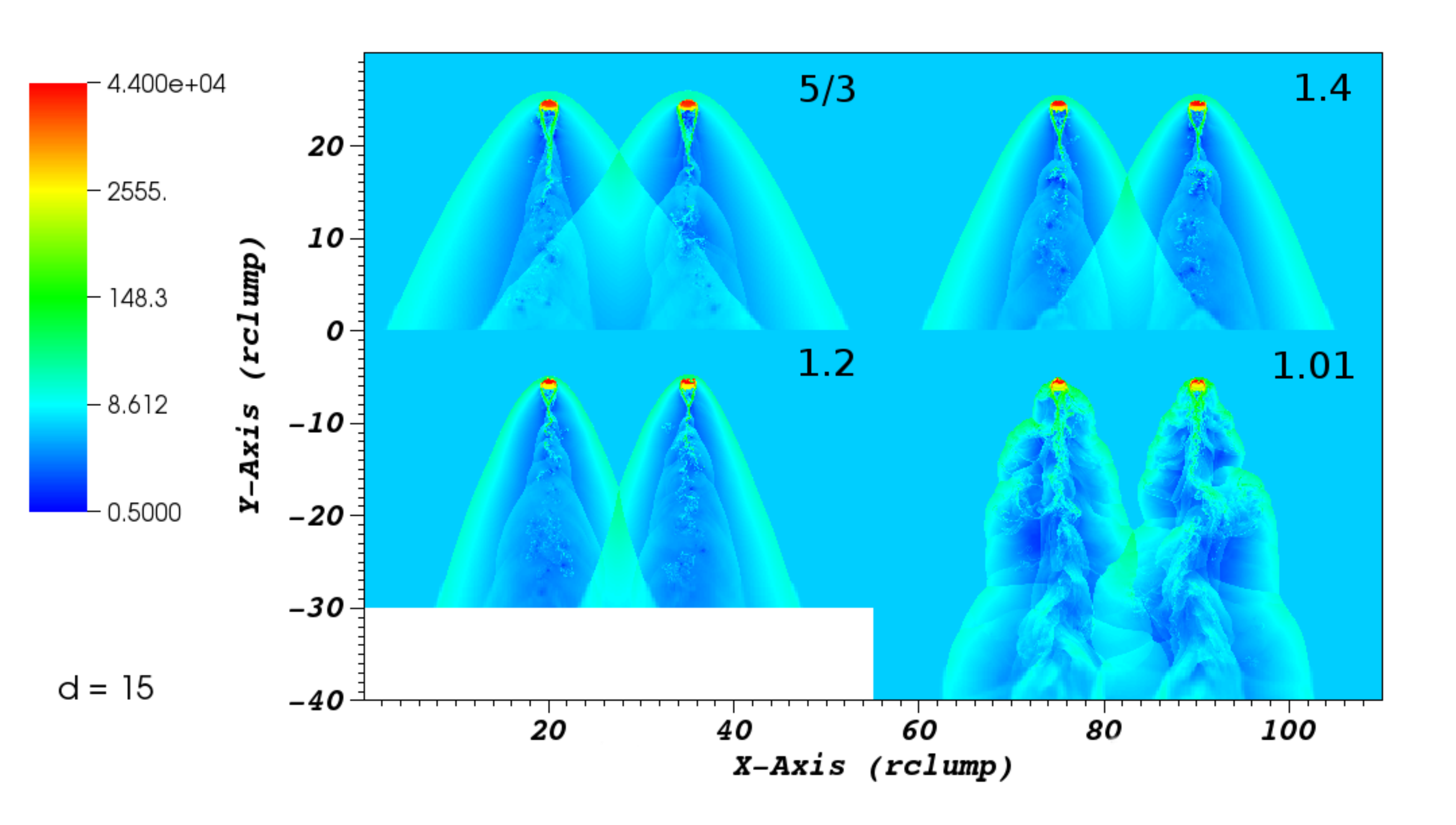}
\caption{Morphology of intersecting bow shocks with a clump separation of 15 $r_{clump}$.
Figure shows number density in units of 1000 cm\textsuperscript{-3} at a simulation time of approximately 75 years.
Each panel is labeled with its corresponding $\gamma$.
None of these models show evidence of a Mach stem.}
\label{fig:d15}
\end{figure}

Figure~\ref{fig:d15} shows density maps for all $\gamma$ value simulations with a separation distance of 15 $r_{clump}$.
Note the strongly unstable nature of the $\gamma = 1.01$ bow shocks.
A very low $\gamma \sim 1$ is often used to simulate isothermal conditions, and hence strong cooling.
Optical observations show evidence of strong cooling in HH objects via [S II] emission maps \cite{Hartigan11}.
Hence the $\gamma = 1.01$ models may be the most physically realistic.
Thus one would expect that if Mach stems form in realistic HH flows they would be short-lived as the angle between bow shocks is constantly changing.
However, as shown by Yirak et al. \cite{Yirak13} Mach stems that have already formed can ``survive'' within a subcritical environment.
Depending on how the angle changes, Mach stem can either grow or be destroyed as the conditions in the flow change.
Very little work has been done on this topic, so we argue that this needs to be explored further in an astrophysical context.

None of the models in Figure~\ref{fig:d15} show evidence of a Mach stem, which is consistent with the theory.
The $d = 20$ cases also showed no evidence of a Mach stem, so we have not shown them here.
Although there are no Mach stems in these models, they each have a high density region in between the reflected shocks.
In 2-D this region appears as a ``cone'' of high density material where the gas has traveled through both the incident bow shock and the reflected shock.
Since this gas is shocked twice, it leads to higher densities and temperatures than just a single bow shock at the same incident angle.

\begin{figure}
\includegraphics[width=\linewidth]{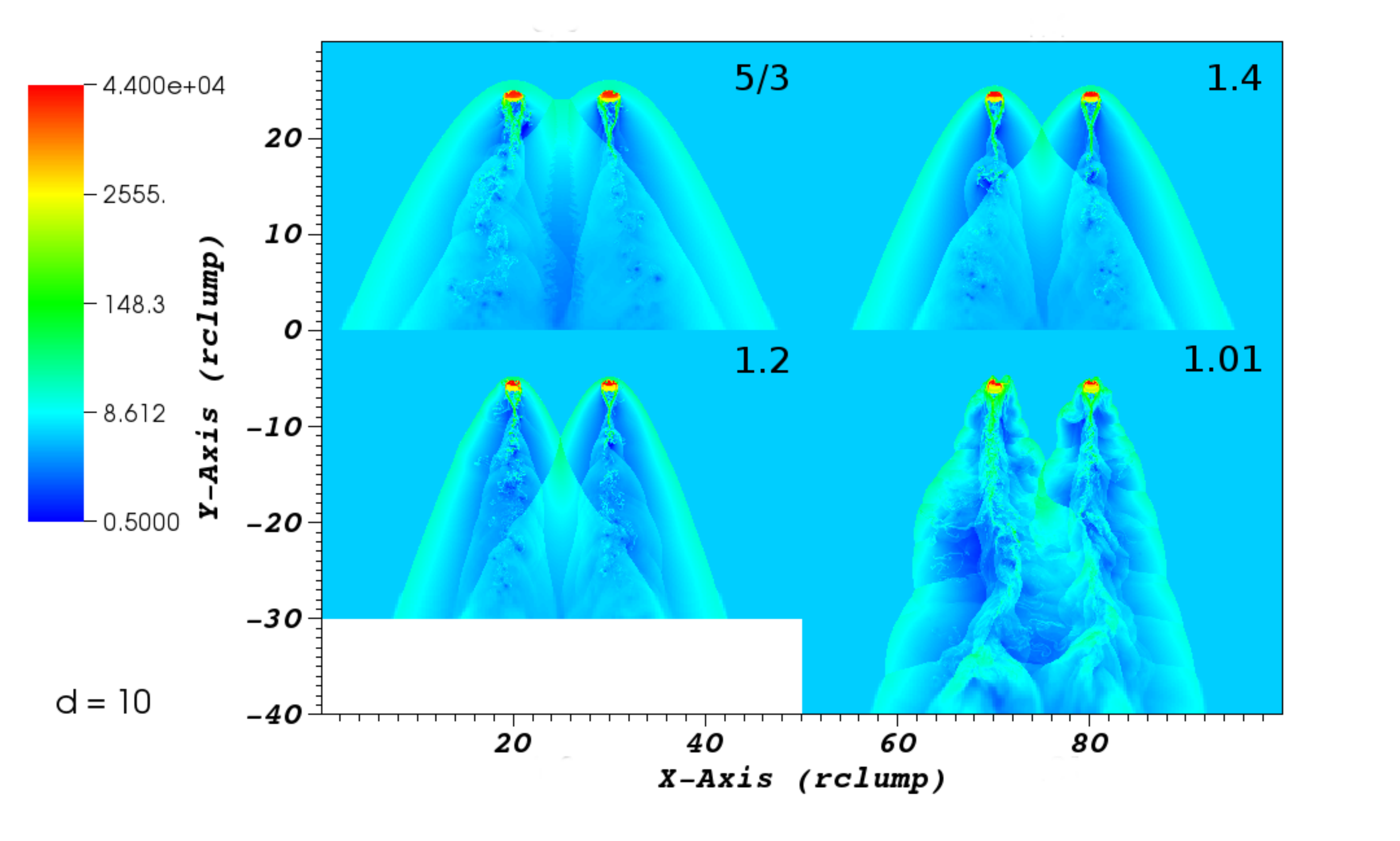}
\caption{Morphology of intersecting bow shocks with a clump separation of 10 $r_{clump}$.
Figure shows number density in units of 1000 cm\textsuperscript{-3} at a simulation time of approximately 75 years.
Each panel is labeled with its corresponding $\gamma$.
The $\gamma = \frac{5}{3}$ model now clearly shows a Mach stem.}
\label{fig:d10}
\end{figure}

\begin{figure}
\includegraphics[width=\linewidth]{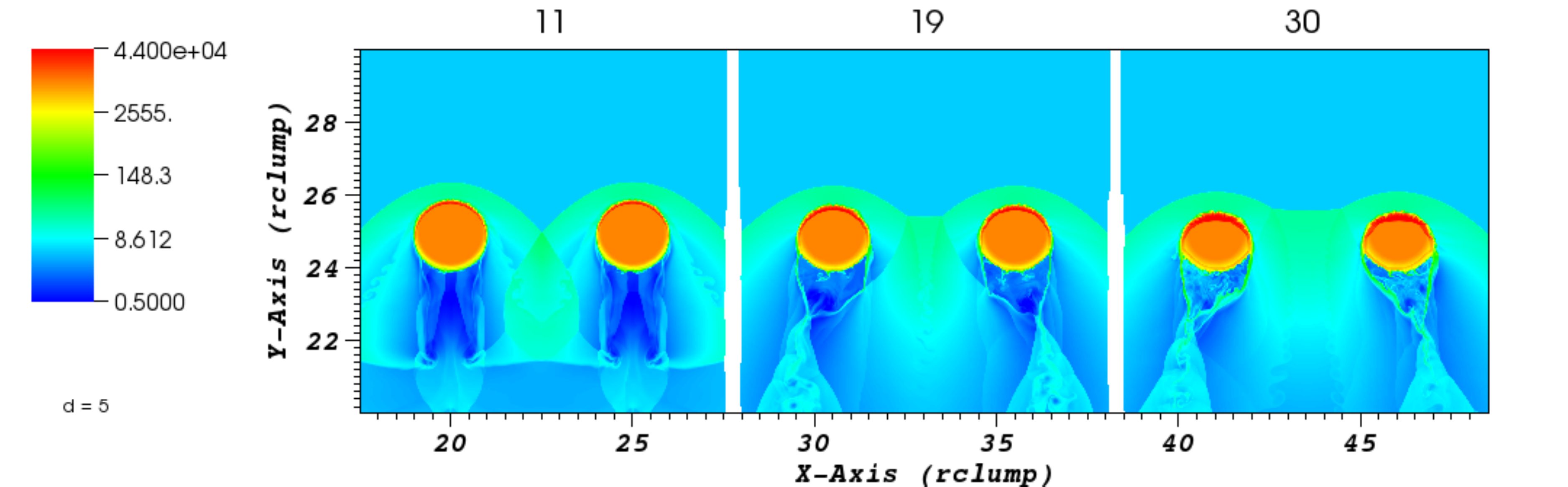}
\caption{Time evolution of the formation of a Mach stem.
This enlarged image shows number density in units of 1000 cm\textsuperscript{-3} for the $\gamma = 1.4$, $d = 5$ model.
Each panel is labeled with its corresponding simulation time in years.}
\label{fig:time}
\end{figure}

Figure~\ref{fig:d10} shows density maps for all cases of $\gamma$ with a separation distance of 10 $r_{clump}$.
The $\gamma = \frac{5}{3}$ model has now clearly formed a Mach stem, and this is consistent with its critical separation distance of 12.5 $r_{clump}$.
Most of the $d = 5$ models are not shown here, but they also agree with the theory.
The $\gamma = \frac{5}{3}$ case still showed a Mach stem, and the $\gamma = 1.4$ case (shown in Figure~\ref{fig:time}) also has a Mach stem.
This newly formed Mach stem in the $\gamma = 1.4$ model is consistent with its critical separation distance of 4.6 $r_{clump}$.

Figure~\ref{fig:time} shows a time series of the $\gamma = 1.4$ model with a separation distance of 5 $r_{clump}$.
In the first frame (at 11 years), we see the bow shocks intersect as usual, forming a pair of incident and reflected shocks.
However, because the intersection angle is above critical, the incident and reflected shocks cannot direct the gas downstream.
This means that the gas has lateral movement which creates an over-dense region in the post-shock ``cone''.
The enhanced density supplies more pressure which forces a Mach stem upstream (shown in frame 2 of Figure~\ref{fig:time}).
In the last frame (at 30 years), the Mach stem has grown even further, and it remains at this size for the remainder of the simulation.

\section{Conclusions}
\label{sec:conc}
Is this paper we have focused on the well-defined theory on Mach reflection and the conditions under which Mach stems form and applied them to astrophysical bow shocks formed by hypersonic clumps.
For strong shocks, the critical angle $\upvarphi_c$ between intersecting comoving bow shocks depends only on $\gamma$.
Based on the shapes of the bow shocks in our simulations, we were able to derive a critical clump separation distance $d_c$.
Smaller $\gamma$ implies a larger critical angle, or shorter critical separation distance.
The formation of Mach stems in our simulations agree with the theoretically derived critical separation distances.

We showed how a Mach stem forms over time with Figure~\ref{fig:time}.
When the angle of intersection is above critical, the intersecting gas creates back pressure in the post-shock region which pushes a Mach stem upstream.
The Mach stem eventually reaches a steady state and stops moving or growing.

Observations in H$\alpha$ show enhanced emission where bow shocks intersect and potentially form Mach stems \cite{Hartigan11}.
It is unclear whether the bright emission is caused by a Mach stem or possibly by a pair of incident and reflected shocks.
More comparisons need to be done on the post-shock conditions in both scenarios.
We know that the answer must depend on the non-equilibrium ionization fraction of the gas.
We hope to answer this question in an upcoming, more extensive paper. 

Values of $\gamma < 1.2$ lead to instabilities along the bow shocks that complicate the problem \cite{Vishniac83}.
Implementing strong non-equilibrum cooling would have a similar effect and make Mach stem studies more challenging.
Included angles would no longer be constant, hence Mach stems could be short-lived flow features.
The observations show that emission features can brighten or dim over short time periods of less than 10 years \cite{Hartigan11}.
While it is possible that these indicate the presence of Mach stems as transient features there could be a wide range of lifetimes.
We did not see any evidence of transient Mach stems in these simulations, and it is likely that 3-D simulations with realistic cooling will be needed to resolve this outstanding issue.

Experiments show that Mach stems can survive below the critical angle, but only for a finite period of time \cite{Yirak13}.
Mach stems need to be studied more thoroughly in the astrophysical context, and this is part of our ongoing research efforts.
In future work, we will explore intersecting bow shocks with different relative velocities and address questions involving 3-D and magnetic field effects as well.

\vspace{7 mm}
\noindent\emph{Acknowledgements.} This work used the Extreme Science and Engineering Discovery Environment (XSEDE), which is supported by National Science Foundation grant number OCI-1053575.
The CIRC at the University of Rochester provided computational resources.
Financial support for this project was provided by the LLE at the University of Rochester, Space Telescope Science Institute grants HST-AR-11251.01-A and HST-AR-12128.01-A; by the National Science Foundation under award AST-0807363; by the Department of Energy under award de-sc0001063.

\section*{References}
\bibliography{mylibrary}
\end{document}